\documentclass[epj,final]{svjour}
\usepackage{graphics}
\usepackage{bm}
\usepackage{graphicx}
\usepackage{color}
\usepackage[usenames,dvipsnames]{xcolor}
\usepackage{epsfig}
\usepackage{longtable}
\usepackage{amsmath}

\def\ket#1{ $ \left\vert  #1   \right\rangle $ }
\def\ketm#1{  \left\vert  #1   \right\rangle   }

\def\bram#1{  \left\langle  #1   \right\vert   }

\def\mem#1#2#3{  \left\langle #1 \left\vert  #2 \right\vert #3 \right\rangle   }

\begin{document}
\title{Bessel beams of two-level atoms \\ driven by a linearly polarized laser field}

\author{A.G. Hayrapetyan\inst{1,2}\thanks{armen@physi.uni-heidelberg.de},
             O. Matula\inst{1,3}, A. Surzhykov\inst{4,5} \and S. Fritzsche\inst{4,5}}

\institute{Physikalisches Institut, Ruprecht-Karls-Universit\"a{}t
Heidelberg,
             D-69120 Heidelberg, Germany
             \and Max-Planck-Institut f\"ur Kernphysik, Postfach 103980,
             D-69029 Heidelberg, Germany
             \and GSI Helmholtzzentrum f\"ur Schwerionenforschung,
             D-64291 Darmstadt, Germany
             \and Helmholtz-Institut Jena, Fr\"obelstieg 3, D-07743 Jena, Germany
             \and Theoretisch-Physikalisches Institut, Friedrich-Schiller-Universit\"at Jena,
             Max-Wien-Platz 1, D-07743 Jena, Germany}

\date{Received: date / Revised version: date}
%
\abstract{We study Bessel beams of two-level atoms that are driven
by a linearly polarized laser field. Starting from the Schr\"odinger
equation, we determine the states of two-level atoms in a plane-wave
field respecting propagation directions both of the atom and the
field. For such laser-driven two-level atoms, we construct Bessel
beams beyond the typical paraxial approximation. We show that the
probability density of these atomic beams obtains a non-trivial,
Bessel-squared-type behavior and can be tuned under the special
choice of the atom and laser parameters, such as the nuclear charge,
atom velocity, laser frequency, and propagation geometry of the atom
and laser beams. Moreover, we spatially and temporally characterize
the beam of hydrogen and selected (neutral) alkali-metal atoms that
carry non-zero orbital angular momentum (OAM). The proposed
spatiotemporal Bessel states (i) are able to describe, in principle,
\textit{twisted} states of any two-level system which is driven by
the radiation field and (ii) have potential applications in atomic
and nuclear processes as well as in quantum communication.
\PACS{{03.65.-w}{Quantum Mechanics} \and
      {03.75.Be}{Atom and neutron optics} \and
      {32.80.Qk}{Coherent control of atomic interactions with photons} \and
      {41.85.Ew}{Particle beam profile, beam intensity} \and
      {42.50.-p}{Quantum Optics}
          } 
} 

\authorrunning{A.G. Hayrapetyan \textit{et al.}}

\titlerunning{Bessel beams of two-level atoms driven by a
                      linearly polarized laser field}

\maketitle
\section{Introduction}
\label{sec:intro}

The discovery of non-diffracting light fields by Durnin \textit{et
al.} \cite{Durnin:87} triggered an interest in studying Bessel beams
of photons \cite{McGloin:05,Jaur:05}. An intriguing property of
these beams is that their intensity does not significantly change
its shape over a propagation distance of a few centimeters
\cite{Tiwari:12,Ismail:12}. The Bessel beams can also have several
unusual features. For instance, they can carry a non-zero OAM and,
thus, make their wavefronts rotate around the propagation axis while
the Poynting vector draws a corkscrew, so called \textit{twisted
photons} \cite{Allen:03-Torres:11}. The simplest object demonstrated
to have such characteristics is the Laguerre-Gaussian beam as
constructed in the seminal paper by Allen \textit{et al.}
\cite{Allen:92}. Twisted photons, moreover, have led to recognizable
advances in optical tweezers \cite{Arlt:01a,Grier:03}, atom trapping
and guiding \cite{Friese-Arlt-Schmid-Liu}, transfer of OAM to a
system of atoms \cite{Tabosa:99-Garces:03}, influence of OAM on beam
shifts \cite{Merano:10-Aiello:12} and in other diverse applications
\cite{Mair-Andersen-Franke-Arnold}.

In analogy with twisted photons, Bliokh \textit{et al.} proposed a
method to construct non-plane-wave solutions of Schr\"odinger
\cite{Bliokh:07} and Dirac \cite{Bliokh:11} equations for a free
electron that exhibits non-zero OAM, called \textit{twisted
electrons}. These \textit{vortex} beams of electrons have been
produced experimentally for an energy of $\sim 200 - 300$ keV only
recently corresponding to the non-relativistic regime
\cite{Uchida:10}. The electron vortex beams have found different
applications as well, such as the magnetic mapping with atomic
resolution in an electron microscope with twisted electrons
\cite{Verbeeck:10-11} and improvement of electron microscopy of
magnetic and biological specimens \cite{McMorran:11}.

Although the photon and electron vortex beams have been quite
extensively investigated, so far the first contribution in exploring
the \textit{atomic} Bessel beams, to the best of our knowledge, has
been illustrated only in \cite{Hayrapetyan:13}. In an effort to
extend the study of photon and electron vortex beams to other type
of beams, we here demonstrate the construction of Bessel beams of
\textit{light} two-level atoms that are resonantly driven by the
plane monochromatic laser beam.

In this work, we investigate Bessel beams of two-level atoms that
are (resonantly) driven by a linearly polarized laser field. For
these beams, we obtain solutions of the Schr\"odinger equation in
terms of the \textit{laser phase} by using an (invariant) approach
as known from the relativistic theory \cite{Berestetskii:82}. We
utilize these solutions in order to construct atomic
\textit{twisted} states that go beyond the typical paraxial
approximation. Furthermore, we examine the spatial and temporal
characteristics of the atomic Bessel beam \textit{profile}. In
particular, detailed calculations are performed especially for the
probability density of hydrogen and light alkali-metal atom beams,
that propagate perpendicular to the propagation direction of the
laser beam. In this \textit{crossed-beam} scenario, we show that the
profile of atomic beams obtain a Bessel-squared-shape that is
modified as the nuclear charge enlarges and can be tuned under the
proper choice of laser parameters, such as the (resonant) frequency
and the electric field strength. In addition, we also exhibit a
possible enhancement of the second maximum in the profile of the
potassium beam should the atom evolve in the field within a
relatively long period, though in a time less than that of atomic
decay.

The paper is organized as follows. In the next section, we define
the twisted states of a freely moving atom and describe their main
properties. In order to see how the laser beam influences the
Bessel-squared-type distribution of the atomic beam profile, in
Subsection \ref{sec:coupling}, we define the classical laser field
and the geometry of the laser and atomic beams propagation, namely
the collinear- and crossed-beam scenarios. For the interaction of a
two-level atom with a linearly polarized monochromatic plane-wave
field, equations are then derived for the atomic probability
amplitudes within the center-of-mass frame of the atom.
\textit{Exact} analytical solutions to these equations are found
within the eikonal (EA), the long-wave (LWA) and the rotating-wave
(RWA) approximations. These solutions are then utilized in
Subsection \ref{subsec:definition-twisted-states-laser-driven-atom}
in order to construct the twisted states of laser-driven two-level
atoms in both, collinear- and crossed-beam scenarios. In Section
\ref{sec:Prob.dens}, we analyze the probability density of these
(Bessel) beams of hydrogen and selected alkali-metal atoms, such as
Li, Na and K, when (resonantly) driven on the $1s \leftrightarrow
2p$, $2s \leftrightarrow 2p$, $3s \leftrightarrow 3p$ and $4s
\leftrightarrow 4p$ atomic transitions, respectively. In particular,
we display and discuss the spatiotemporal characteristics of these
beams in the crossed-beam scenario. Finally, a few conclusions  and
proposals are drawn in Section \ref{Conclusion} and a detailed
derivation of the solution to the Schr\"odinger equation (with
respecting propagation directions both of the atomic and the laser
beams) is illustrated in Appendix \ref{app:derivation}.

\section{Twisted states of free atoms}
\label{sec:definition-twisted-states-free-atom}

\begin{figure}
\center
\includegraphics[width=0.48\textwidth]{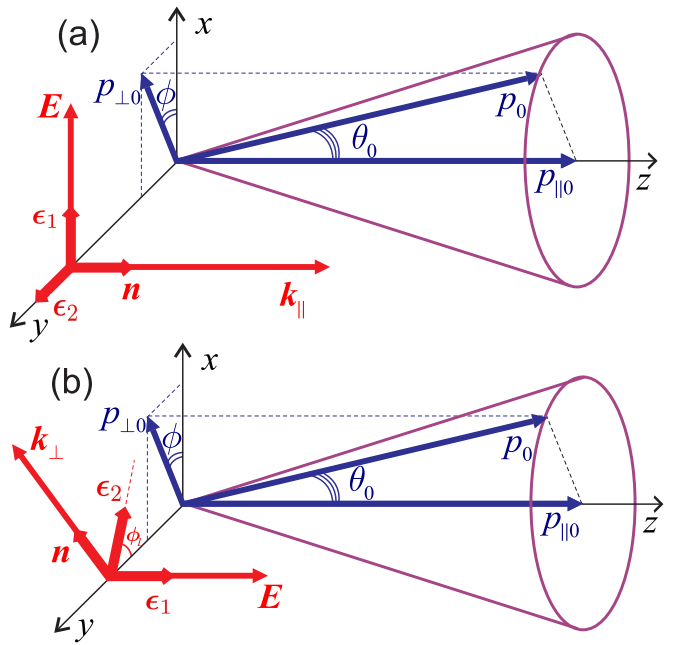}
\caption{(Color online). Collinear-beam (a) and crossed-beam (b)
scenarios for the interposition of linearly polarized laser and
atomic Bessel beams. Red sketches indicate directions of the field
polarization and the wave vector. Blue sketches indicate the
momentum distribution of the atomic Bessel beam. A twisted two-level
atom driven by the laser field is a superposition of the
orthonormalized waves (\ref{solution1}) (a) or (\ref{solution2}) (b)
with a fixed conical momentum spread $p_0 = const$, polar angle
$\theta_0$ and the azimuthal phase factor $e^{i \ell \phi}$.}
\label{fig.1}
\end{figure}

A Bessel beam of any quantum particle is defined as a
(\textit{twisted}) state with its well defined energy ${\cal E}$,
longitudinal momentum $p_{||}$, absolute value of the transverse
momentum $p_{\bot}$ as well as the projection $\hbar \ell$ of the
OAM on the propagation axis
\cite{Allen:03-Torres:11,Bliokh:11,Jentschura:11}. In accordance to
this definition, the spectrum of such states can be represented in
the form
\begin{eqnarray}
\label{Bessel}
    {\tilde{\psi}}_{\ell} \left( \bm p \right)
    & = &
    \delta ( {\cal E} - {\cal E}_0)
    \delta \left( p_{\bot} - p_{\bot 0} \right)
    \delta ( p_{|| } - p_{ || 0})
    \frac{e^{i \ell \phi}}{2\pi i^\ell p_{\bot 0}} \quad
\end{eqnarray}
that contains a vortex phase dependency $e^{i \ell \phi}$, and where
$\ell$ is an integer. Such spectrum means that the momentum of the
atom is distributed over some cone with slant length $p_0 =
\sqrt{p_{|| 0}^2 + p_{\bot 0}^2} = const$ and fixed polar angle
$\theta_0$ with regard to the $z$-axis [cf.\ Figure \ref{fig.1}, the
blue sketch]. The `size' of this cone is defined such as the
longitudinal and transverse components of the beam momentum are
equal to $p_{|| 0} = p_0 \cos{\theta_0}$ and $p_{\bot 0} = p_0
\sin{\theta_0}$, respectively. Owing to the conical symmetry of the
momentum distribution, we shall use cylindrical coordinates $\bm p =
( p_{\bot}, \phi, p_{||}) =  \left( p \sin \theta, \phi, p \cos
\theta \right)$ in momentum space and construct a Bessel-type
solution of the Schr\"odinger (free wave) equation
\begin{eqnarray}
\label{integral}
    \psi_\ell \left(\bm r , t \right) & = &
    \int{ {\tilde{\psi}}_{\ell}
          \left( \bm p \right)  \, \psi^{\scriptscriptstyle PW} \! \!\left(\bm r , t \right)
          p_{\bot}d p_{\bot} d \phi \, d p_{||}} \,
 \end{eqnarray}
as a superposition of plane waves
\begin{eqnarray}
\label{plane-wave-atom}
    \psi^{\scriptscriptstyle PW}\!\! \left(\bm r , t \right) =
    e^{\frac{i}{\hbar} \left( \bm p \cdot \bm r - {\cal E} t \right)}
\end{eqnarray}
over the spectrum (\ref{Bessel}) of such a cone.

\begin{figure}
\includegraphics[width=0.48\textwidth]{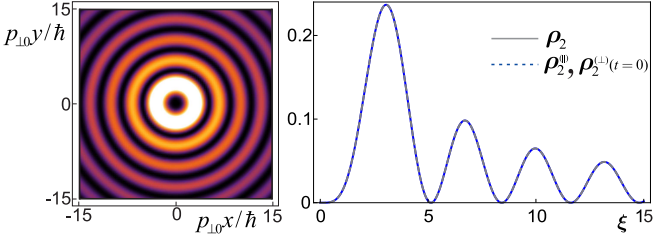}
\caption{(Color online). Distributions of probability density (in
arbitrary units) over the dimensionless transverse coordinate for
free (gray solid curve) and laser-driven (blue dashed curve) beams
of twisted atoms with OAM $\hbar \ell = 2 \hbar$. For the
crossed-beam scenario, the probability density
$\rho_2^{\scriptscriptstyle (\bot)}$ is depicted at $t = 0$ when the
field is switched off. The distribution is shown by the variation of
colors from black to white within the `sunset' scale (left panel),
where black and white correspond to the minimum and maximum values
of the probability density.} \label{fig.2}
\end{figure}

Integration (\ref{integral}) can be readily performed if we take
into account the cylindrical symmetry of the atomic beam propagation
and use the cylindrical coordinates in the position space as well,
$\bm r = \left( r, \varphi, z \right)$. These coordinates, along
with the cylindrical coordinates in momentum space, enable one to
re-write the scalar product as
\begin{eqnarray*}
    \bm p \cdot \bm r \,=\, p_\bot r \cos \left( \phi-\varphi \right) +
    p_{||} z
\end{eqnarray*}
in the plane wave (\ref{plane-wave-atom}) and, therefore, to exploit
the integral representation
\begin{eqnarray}
\label{int-repr}
   \int_{0}^{2 \pi}
   \!\! \mathrm{d}  \phi \, e^{ i \ell \phi } e^{ i \xi \cos ( \phi -
   \varphi)} = 2 \pi i^\ell  e^{i \ell \varphi} J_\ell \left( \xi \right)
\end{eqnarray}
of the Bessel function \cite{Gradshtein:00}. Direct integration
simply leads to the final form of the twisted state of a free atom
\begin{eqnarray}
\label{free-twisted-state}
   \psi_\ell \left(\bm r , t \right) & = &
               e^{\frac{i}{\hbar} (p_{|| 0} z - {\cal E}_0 t )}
               e^{i \ell \varphi} J_\ell \left( \xi \right) \, ,
\end{eqnarray}
where $\xi = p_{\bot 0} r/\hbar$ is a dimensionless transverse
coordinate  which characterizes the \textit{width} of the beam. As
seen from Eq.~(\ref{free-twisted-state}), the state $\psi_\ell
\left(\bm r , t \right)$ represents an atomic beam that propagates
freely along the $z$-direction, $e^{i p_{|| 0} z/ \hbar}$, and has
the profile
\begin{eqnarray}
\label{free-profile}
         \rho_\ell \equiv \left| \psi_\ell \left(\bm r , t \right) \right|^2
         = J_\ell^2 \left( \xi \right) \, ,
\end{eqnarray}
a Bessel-squared-shape in the radial dimension (as also illustrated
in Figure \ref{fig.2}). The vortex phase factor $e^{i \ell \varphi}$
ensures that the twisted state (\ref{free-twisted-state}) is an
eigenstate of the $z$-component of OAM operator $\hat{\ell}_z = - i
\hbar \partial / \partial \varphi$, and, thus, is responsible for
non-zero OAM $\hbar \ell$ on the propagation axis. In addition, one
should stress that the states (\ref{free-twisted-state}) are
orthogonal and can be normalized if the integration is carried out,
for instance, over a large, but finite cylindrical volume. For our
further analysis, however, the normalization is not of crucial
interest.

We have shown a simple, but very significant procedure how one can
create twisted states of a free particle (e.g. atom). In our later
study, we are interested in how these twisted states are affected by
an external influence, such as a driving laser field. For this
purpose, we will initially prepare a two-level atom in the upper
state and then \textit{switch on} the laser field. Once this atom is
coherently driven by the field, we will apply the same approach in
order to construct Bessel beams of such atoms and to investigate
their spatiotemporal characteristics.

\section{Twisted states of laser-driven atoms}
\label{sec:definition-twisted-states-laser-driven-atom}

In the previous section, we have shown a simple way of construction
of free atomic Bessel beams. In this section, we are interested in
the dynamics of these beams when they are (coherently) driven by the
field of a plane mono-chromatic laser wave. To this end, in
subsection \ref{sec:coupling}, we examine the coherent coupling of a
two-level atom with the linearly polarized radiation field, and
then, in Subsection
\ref{subsec:definition-twisted-states-laser-driven-atom}, we
construct the twisted states of this laser-driven atom by employing
the (Bessel) spectrum (\ref{Bessel}).

\subsection{Semi-classical coupling of two-level atoms to a laser field}
\label{sec:coupling}

The coherent interaction of a two-level atom with the radiation
field has been explored since the early days of quantum optics
\cite{Scully/Zubairy:01,Allen/Eberly:87-Foot:05,Cohen-Tannoudji:11}.
Nowadays, this \textit{atomic coherence} is known as an effective
tool for achieving control about (atomic and molecular) samples
\cite{Kozlov-Champ-Card} and it can be utilized to examine various
effects with atomic beams, such as generation of entanglement
\cite{Duan:00}, investigation of the optical force \cite{Noh-Kumar}
or exhibition of vortices \cite{Andrelczyk-Helseth-Liu} in beams. In
order to examine how this coherent control can be exploited
\textit{also} for tuning and guiding the atomic Bessel beams, in the
following, we shall describe the coherent coupling of an atom to the
laser field in the meantime respecting the propagation directions of
both the atom and laser beams.

To describe the interaction of a two-level atom with a
\textit{classical} radiation field and to start our derivations, let
us first characterize the field and the interaction (Hamiltonian)
between the atom and this field. We hereby assume that the atom has
the (rest) mass $m$, a dipole moment $\bm{d}$, and that it moves
with constant momentum $\bm{p}$ along some \textit{given} direction.

\subsubsection{Characterization of the classical field}
\label{subsubsec:field}

We consider the classical electric field of laser beam as a
(monochromatic) plane wave
\begin{eqnarray}
\label{plane-wave}
    \bm{E} & = &
    \bm{\varepsilon} \, e^{i \left(\bm k \cdot \bm r - \omega t\right)} \, ,
\end{eqnarray}
with constant electric-field amplitude $\bm \varepsilon $, frequency
$\omega$ and wave vector $\bm k \,\equiv\, k \,\bm{n}$, which
satisfies Maxwell's equations, and where the unit vector $\bm{n}$
defines the propagation direction of the wave \cite{Jackson:01}.
While, in general, $\bm{E}$ and $\bm{\varepsilon}$ are both
complex-valued vectors in Eq.~(\ref{plane-wave}), the physically
relevant \textit{electric} field is given by the real part, $\Re
(\bm{E})$, and is for transverse waves always perpendicular to its
propagation, i.e.\ $\bm{n} \cdot\, \bm{\varepsilon} \,=\, 0$. For
our further discussion, moreover, it is useful to introduce a set of
real and mutually orthogonal unit vectors $\left(\bm{\epsilon}_1 \,
, \bm{\epsilon}_2 \, , \bm n \right)$, and to re-write the field
amplitude in terms of these vectors as
\begin{eqnarray}
\label{amplitude}
    \bm \varepsilon & = & \bm{\epsilon}_1 \,\varepsilon_0 \:+\:
                          \bm{\epsilon}_2 \,\varepsilon_0^\prime \, ,
\end{eqnarray}
with the two (complex) constants $\varepsilon_0$ and
$\varepsilon_0^\prime$.

Most generally, the laser field amplitude $\bm{\varepsilon}$ in
Eq.~(\ref{plane-wave}) describes an \textit{elliptically} polarized
plane  wave; as a special case, this definition includes
\textit{linearly} polarized waves if the complex constants
$\varepsilon_0$ and $\varepsilon_0^\prime$ fulfill proper relations.
We choose $\varepsilon_0^\prime =0$ which corresponds to a linearly
polarized field in $\bm{\epsilon}_1$ direction
\begin{eqnarray*}
    \bm{E} \left(\bm{r}, t \right) & = &
    \bm{\epsilon}_1 \, \varepsilon_0 \,
    e^{i \left( \bm{k} \cdot \bm{r} - \omega t \right)}
\end{eqnarray*}
with the (real) amplitude $\varepsilon_0$. Therefore, for this wave
the relevant electric field is given by
\begin{eqnarray}
\label{lin-pol}
    \bm E  & = & \bm{\epsilon}_1 \varepsilon_0 \cos \zeta
    = \bm \varepsilon \cos \zeta \, ,
\end{eqnarray}
where $\zeta \,\equiv\,\bm{k}\cdot\bm{r}-\omega t$ is the
\textit{phase} of the plane wave independent of its particular
polarization properties.  We here should stress that the field is
evaluated at the center-of-mass position $\bm r \equiv \left( x, y,
z \right)$ of the atom.

The field and the atom propagation directions, in principle, can be
chosen arbitrarily which will form the geometry of the ``atom +
laser" system. To describe the linearly polarized laser field we
specify this geometry already in this subsection. We examine two
scenarios that depend on the interposition of the propagation
directions of laser and atomic beams. These are the so-called
\textit{collinear-} and \textit{crossed-beam} scenarios for which
the laser and the atomic beam propagation directions correspondingly
are parallel or perpendicular to each other [cf.\ Figure
\ref{fig.1}]. As mentioned above, the $z$-axis is chosen along the
propagation direction of the atomic beam. In collinear-beam
scenario, moreover, we chose the $x$- and $z$-axes directed along
the polarization $\bm \epsilon_1$ and laser beam propagation,
respectively. In contrast to this, in crossed-beam scenario, we
restrict ourselves with the polarization along $z$-axis and consider
propagation in $x-y$-plane, where the $x$-axis is declined from the
laser propagation direction under $\phi_L$ angle.

With the distinction of these collinear- and crossed-beam scenarios
the laser field (\ref{lin-pol}) and the phase $\zeta$ obtain the
form
\begin{eqnarray}
\label{lin-pol-col}
    \bm E^{{\scriptscriptstyle (||)}} & = & \left( \varepsilon \cos \zeta^{{\scriptscriptstyle (||)}} , 0 , 0 \right)  , \;
    \zeta^{{\scriptscriptstyle (||)}} = k_{||} z - \omega t \, ,
\\[0.3cm]
\label{lin-pol-cros}
    \bm E^{{\scriptscriptstyle (\bot)}} & = & \left( 0 , 0 , \varepsilon \cos \zeta^{{\scriptscriptstyle (\bot)}} \right) , \;
    \zeta^{{\scriptscriptstyle (\bot)}}= k_{\bot} r \cos \left( \phi_L - \varphi \right)  - \omega t \, , \quad \quad
\end{eqnarray}
respectively, with a transverse coordinate $r \equiv \sqrt{x^2+y^2}$
of the atomic center-of-mass. In the following, we define the
Hamiltonian of our system and determine the atomic states that are
driven by the laser field (\ref{lin-pol-col}) and
(\ref{lin-pol-cros}).

\subsubsection{Solution of the Schr\"odinger equation in the center-of-mass frame of the atom}
\label{subsubsec:atom-lin.field}

In the semi-classical theory, the quantum dynamics of the atom is
driven by the external field but this ``motion'' does \textit{not}
re-act back upon the field as this would require a quantization of
the electromagnetic field \cite{Scully/Zubairy:01}. Therefore, the
Hamiltonian of an atom in the (classical) field takes the form
\begin{eqnarray}
\label{Ham}
   H & = & \frac{\hat{p}^2}{2m} + H_{\rm atom} + H_{\rm int},
\end{eqnarray}
where $\frac{\hat{p}^2}{2m}$ denotes the kinetic energy (operator)
as associated with the center-of-mass motion and
\begin{eqnarray}
\label{at.Ham}
   H_{\rm atom} & = &  E_a \ketm{a}\bram{a} + E_b \ketm{b}\bram{b}
\end{eqnarray}
refers to the internal motion of the atom, whilst the interaction of
the atom and the field is taken (as usual) in \textit{minimal}
coupling
\begin{eqnarray*}
   H_{\rm int} & = & -\bm{d} \cdot \bm{E}
\end{eqnarray*}
in the LWA \cite{Cohen-Tannoudji:11}. The two state vectors
$\ketm{a}$ and $\ketm{b}$ in expression (\ref{at.Ham}) denote the
upper and lower states of the atom and are supposed to be
eigenstates of the atomic Hamiltonian
\begin{eqnarray*}
   H_{\rm atom} \ketm{a} & = & E_a \ketm{a}\,  \\
   H_{\rm atom} \ketm{b} & = & E_b \ketm{b}\, ,
\end{eqnarray*}
and where $E_a$ and $E_b$ are the energies of upper and lower
states, respectively.

Making use of the completeness $\ketm{a}\bram{a} + \ketm{b}\bram{b}
\,=\, 1$ for a two-level system, the interaction Hamiltonian of the
atom with a linearly-polarized wave can be written in the form
\begin{eqnarray}
\label{int.Ham1}
   H_{int}^{{\scriptscriptstyle (||)}} & = & -\hbar \Omega_{R_x} \!
                  \left( e^{i\phi_{d_x}}   \ketm{b}\bram{a}
                         +e^{-i\phi_{d_x}} \ketm{a}\bram{b}
                  \right) \! \cos\zeta^{{\scriptscriptstyle (||)}}  , \quad
  \\
\label{int.Ham2}
   H_{int}^{{\scriptscriptstyle (\bot)}} & = & -\hbar \Omega_{R_z} \!
                  \left( e^{i\phi_{d_z}}   \ketm{b}\bram{a}
                         +e^{-i\phi_{d_z}} \ketm{a}\bram{b}
                  \right) \! \cos\zeta^{{\scriptscriptstyle (\bot)}} \quad \quad
\end{eqnarray}
for collinear- and crossed-beam scenarios, respectively. In
accordance to these scenarios, here $\Omega_{R_x} = \left|e\mem{b}{
\tilde{x} }{a}\right|\varepsilon_0/\hbar $ and $\Omega_{R_z} =
\left|e\mem{b}{ \tilde{z} }{a}\right|\varepsilon_0/\hbar $ denote
the Rabi frequencies that describe the coupling strength of the atom
either with the $x$- (\ref{lin-pol-col}) or $z$-polarized
(\ref{lin-pol-cros}) field, respectively. With this distinction for
the Rabi frequencies, the exponentials $\phi_{d_x}$ and $\phi_{d_z}$
denote the phases of the dipole matrix elements: $e\mem{b}{u}{a}
=|e\mem{b}{u}{a}|e^{i\phi_{d_u}}$ where $u = \tilde{x},\tilde{z}$.
Here both $\tilde x$ and $\tilde z$ refer to so-called internal
variables of the atom and are the electron's relative coordinates
with regard to the nucleus \cite{Cohen-Tannoudji:11}. The Rabi
frequencies, moreover, arise naturally as non-diagonal matrix
elements of the atom dipole moment if we express the interaction
Hamiltonian in the $\left\{ \ketm{a} , \ketm{b} \right\}$ basis.
This, in turn, makes the interaction Hamiltonian dependent on atomic
states $\ketm{a}$, $\ketm{b}$ and the center-of-mass coordinate of
the atom \footnote{Hereinafter, for notational convenience we drop
the ``tilde" of the indexes $x$ and $y$ at the Rabi frequencies and
the exponentials of the dipole matrix elements. This will not cause
a confusion since already at this step the dependence on the
electron's relative coordinate is eliminated in the interaction
Hamiltonian (\ref{int.Ham1})-(\ref{int.Ham2}).}.

For such an effective factorization of the interaction Hamiltonian,
and as described in many texts before
\cite{Scully/Zubairy:01,Allen/Eberly:87-Foot:05}, the two-level atom
undergoes Rabi oscillations between its lower and upper states with
frequency $\Omega_{R_x}$ or $\Omega_{R_z}$, quite analogue to a
spin-$1/2$ system in an oscillating magnetic field \cite{Rabi:37}.
In contrast to the standard derivation (see, for example,
Ref.~\cite{Scully/Zubairy:01}), however, the interaction Hamiltonian
(\ref{int.Ham1})-(\ref{int.Ham2}) now also depends on the phase
$\zeta^{\scriptscriptstyle (||)}$ and $\zeta^{\scriptscriptstyle
(\bot)}$ of the radiation field to account for the time-
\textit{and} space-dependency of the atom-laser interaction. In
addition, some time ago an approximate technique, i.e. expansion of
the exponent (\ref{plane-wave}) in the interaction Hamiltonian, has
been applied to the helium and hydrogen atoms in order to examine
their coupling to an intense laser field~\cite{Meharg:05}.

To explore the time evolution of the atom, let us search for
solutions of the (time-dependent) Schr\"o{}dinger equation
\begin{eqnarray}
\label{Sch.eq}
  H \psi & = & i\hbar \frac{\partial\psi}{\partial t}  \, .
\end{eqnarray}
If we utilize the (effective) factorization of the interaction
Hamiltonian we can use the ansatz
\begin{eqnarray}
\label{psi}
   \psi \left(\bm r , t \right)
   & = & e^{\frac{i}{\hbar}\left( \bm{p} \cdot \bm{r} - {\cal E} t \right)}
         \left( \psi_a (\zeta)\, \ketm{a} + \psi_b (\zeta)\, \ketm{b}
     \right) \, ,
\end{eqnarray}
as a solution of the Schr\"odinger equation. Here $\zeta \equiv
\zeta^{\scriptscriptstyle (||)}$, $\zeta^{\scriptscriptstyle
(\bot)}$ is the laser phase for collinear- or crossed-beam
scenarios, respectively. The constant (non-relativistic) momentum
$\bm{p} \equiv m \bm v$ of the center-of-mass of atom gives hereby
rise to the space-dependent ``translation factor''
$e^{\frac{i}{\hbar}\bm{p}\bm{r}}$ to account for its overall motion
with energy ${\cal E}$ within the given coordinates. A similar
ansatz has been applied in \cite{Zeng:01} in order to investigate
Stark splitting of a three-level atom.

We give the detailed derivation of the solution of the Schr\"odinger
equation (\ref{Sch.eq}) in Appendix \ref{app:derivation} where the
physical assumptions we consider are the following. We prepare the
atom initially in the upper state [cf.\ Eq.~(\ref{cond-for-lin})]
and then employ the typical, \textit{eikonal} and \textit{rotating
wave} approximations, for which the atom rest energy is much larger
than the photon energy (EA) that, in turn, is resonant to the atomic
transition energy (RWA, see also Eq.~(\ref{res})). For light
two-level atoms, both the EA and RWA are valid with high accuracy
\cite{Scully/Zubairy:01}. Under these physically relevant
conditions, we obtain analytically exact solution
\begin{eqnarray}
\nonumber
   \psi^{\scriptscriptstyle (||)} \left(\bm r , t \right)
   & = & e^{\frac{i}{\hbar}\left( \bm{p} \cdot \bm{r} - {\cal E} t \right)} \:
         \left( e^{-i \alpha^{\scriptscriptstyle (||)} \zeta^{\scriptscriptstyle (||)}}
         \cos \! \frac{\Omega^{\scriptscriptstyle (||)} \zeta^{\scriptscriptstyle (||)}}{2} \ketm{a}
         \right.
  \\
\label{solution1}
    & + &  \left.
                i e^{i \phi_{d_x}} e^{-i \beta^{\scriptscriptstyle (||)} \zeta^{\scriptscriptstyle (||)}} \sin \!
                \frac{\Omega^{\scriptscriptstyle (||)} \zeta^{\scriptscriptstyle (||)}}{2} \ketm{b}
               \right)
\end{eqnarray}
that represents the state of the laser-driven two-level atom for
collinear-beam scenario with some reduced quantities [cf.\
Eq.~(\ref{reduced-quant})]
\begin{eqnarray}
\label{reduced-quant-col-beam}
   \left\{ \alpha^{\scriptscriptstyle (||)} \, , \, \beta^{\scriptscriptstyle (||)} \, , \,
   \Omega^{\scriptscriptstyle (||)} \right\}  \equiv
   \frac{\left\{ E_a \, , \, E_b \, , \, \Omega_{R_x} \right\}} {\hbar\left( k v_{||}-\omega\right)} \, .
\end{eqnarray}
Whereas, in crossed-beam scenario, the atomic state is given by
\begin{eqnarray}
\nonumber
   \psi^{\scriptscriptstyle (\bot)} \left(\bm r , t \right)
   & = & e^{\frac{i}{\hbar}\left(\bm{p} \cdot \bm{r}- {\cal E} t \right)} \:
         \left( e^{-i \alpha^{{\scriptscriptstyle (\bot)}} \zeta^{{\scriptscriptstyle (\bot)}}}
         \cos \! \frac{\Omega^{{\scriptscriptstyle (\bot)}} \zeta^{{\scriptscriptstyle (\bot)}}}{2} \ketm{a}
         \right.
  \\
\label{solution2}
    & + &  \left.
                i e^{i \phi_{d_z}} e^{-i \beta^{{\scriptscriptstyle (\bot)}} \zeta^{{\scriptscriptstyle (\bot)}}} \sin \!
                \frac{\Omega^{{\scriptscriptstyle (\bot)}} \zeta^{{\scriptscriptstyle (\bot)}}}{2} \ketm{b}
               \right) \,
\end{eqnarray}
with reduced quantities
\begin{eqnarray}
\label{reduced-quant-cros-beam}
  \hbar \left\{ \alpha^{{\scriptscriptstyle (\bot)}} \, , \, \beta^{{\scriptscriptstyle (\bot)}} \, , \,
  \Omega^{{\scriptscriptstyle (\bot)}} \right\}  \equiv
   \frac{\left\{ E_a \, , \, E_b \, , \, \Omega_{R_z} \right\}} {k v_{\bot} \cos ( \phi_L -\phi) -\omega} \, .
\end{eqnarray}
In Eqs.\ (\ref{solution1})-(\ref{reduced-quant-cros-beam}), $p^2
\,=\, p_{||}^2 + p_{\bot}^2$ is the squared total momentum of the
atom (with its longitudinal $p_{||}$ and transverse $p_{\bot}$
components and the atomic velocities $v_{||} = p_{||}/m$ and
$v_{\bot} = p_{\bot}/m$, respectively). The \textit{orthonormalized}
solutions (\ref{solution1}) and (\ref{solution2}) represent a
superposition of the upper $\ketm{a}$ and lower $\ketm{b}$ states
with $\zeta^{\scriptscriptstyle (||)}$- or
$\zeta^{\scriptscriptstyle (\bot)}$-dependent coefficients. The
`translation' factor $e^{\frac{i}{\hbar}\left(\bm{p} \cdot \bm{r}-
{\cal E} t \right)}$, as mentioned above, describes the motion of
the atom as a whole with momentum vector $\bm{p}$ along some
(chosen) direction and energy ${\cal E}$.

So far, the focus of this subsection was placed on determining
solutions of the Schr\"odinger equation (\ref{Sch.eq}) for linearly
polarized field (\ref{lin-pol-col}) or (\ref{lin-pol-cros}). In the
next subsection, we will utilize the \textit{explicit} $\bm
r$-dependency of (exact) solutions (\ref{solution1}) and
(\ref{solution2}) and exploit them in order to construct a
\textit{Bessel beam} of laser-driven two-level atoms.

\subsection{Twisted states of laser-driven two-level atoms}
\label{subsec:definition-twisted-states-laser-driven-atom}

In the previous subsection, we have analyzed the time- and
space-dependent interaction of two-level atoms with a linearly
polarized laser field and have built the states (\ref{solution1})
and (\ref{solution2}) which describe the \textit{spatiotemporal}
dynamics of laser-driven atoms. In the following, we shall utilize
these states in order to construct atomic Bessel beams which carry a
non-zero OAM and meanwhile do not diffract along the propagation
direction.

The wave function of the laser-driven two-level atom with the
projection $\hbar \ell$ of the OAM on the (atomic beam) propagation
axis can be constructed as a superposition of orthonormalized
solutions (\ref{solution1}) and (\ref{solution2}) of the
time-dependent Schr\"odinger equation (\ref{Sch.eq})
\begin{eqnarray}
\label{Integral}
    \Psi_{\ell}\left(\bm r , t \right) & = &
    \int{ {\tilde{\psi}}_{\ell}
          \left( \bm p \right)  \, \psi \left(\bm r , t \right)
          p_{\bot}d p_{\bot} d \phi \, d p_{||}} \,
 \end{eqnarray}
over the \textit{monoenergetic} cone (\ref{Bessel}). This step in
the construction of \textit{twisted} atomic beams is analogue to the
use of twisted photons \cite{Allen:03-Torres:11} and electrons
\cite{Bliokh:11}.

\subsubsection{Collinear-beam scenario}
\label{sec:coll-beam}

Let us start with the collinear-beam scenario for which the
integration (\ref{Integral}) can be carried out easily in a similar
way as for the twisted state (\ref{free-twisted-state}) of a free
atom. Indeed, by performing the same steps, i.e. substituting the
state (\ref{solution1}) into Eq. (\ref{Integral}) and making use of
the integral representation (\ref{int-repr}) of Bessel function, we
obtain a \textit{simple} form for the twisted state of a
laser-driven two-level atom
\begin{eqnarray}
\nonumber
   \Psi_{\ell}^{\scriptscriptstyle (||)} & = &
               e^{\frac{i}{\hbar} (p_{|| 0} z - {\cal E}_0 t )}
               e^{i \ell \varphi} J_\ell \left( \xi \right) \,
               \left(
        e^{-i \alpha_{0}^{\scriptscriptstyle (||)} \zeta^{\scriptscriptstyle (||)}}
        \cos \frac{\Omega_{0}^{\scriptscriptstyle (||)} \zeta^{\scriptscriptstyle (||)}}{2}
        \, \ketm{a}
               \right.
\\
\label{beam-col}
   & + &
   \left.
        i e^{i\phi_{d_x}} e^{-i \beta_{0}^{\scriptscriptstyle (||)} \zeta^{\scriptscriptstyle (||)}}
        \sin \frac{\Omega_{0}^{\scriptscriptstyle (||)} \zeta^{\scriptscriptstyle (||)}}{2} \, \ketm{b}
   \right) \, .
\end{eqnarray}
Here the dimensionless transverse coordinate $\xi = p_{\bot 0}
r/\hbar$ describes the width of the beam, and the reduced quantities
$\alpha_{0}^{\scriptscriptstyle (||)}$,
$\beta_{0}^{\scriptscriptstyle (||)}$ and
$\Omega_{0}^{\scriptscriptstyle (||)}$ are taken on the cone
`surface' $p = p_0 $ [cf.\ Eq.~(\ref{reduced-quant-col-beam}) and
Figure \ref{fig.1}a].

The state (\ref{beam-col}) is the eigenstates of the $z$-component
of OAM operator $\hat{\ell}_z$ due to the presence of the vortex
phase factor $e^{i \ell \varphi}$. Hence, apart from the free
propagation along the $z$-direction, $e^{i p_{|| 0} z/ \hbar}$,  and
the Bessel-dependency on the (dimensionless) transverse coordinate
the twisted state $\Psi_{\ell}^{\scriptscriptstyle (||)} $ also
carries a well-defined OAM $\hbar \ell$ quite similar to the scalar
\cite{Bliokh:07,Schattschneider:11} and spin-dependent
\cite{Bliokh:11} electron Bessel beams.

It is obvious that when the field is switched off, i.e.
$\Omega_{0}^{\scriptscriptstyle (||)} = \zeta^{\scriptscriptstyle
(||)} = 0$, the free twisted state (\ref{free-twisted-state}) is
recovered. The presence of the field causes only Rabi flopping of
(already) twisted two-level atoms between the upper $\ketm{a}$ and
lower $\ketm{b}$ states and meantime does \textit{not} influence the
Bessel-squared-shape of the beam profile. Alluding to Section
\ref{sec:Prob.dens}, let us note that the absence of the field's
impact on the beam profile is quite expected since the laser phase
$\zeta^{\scriptscriptstyle (||)} = k_{||} z - \omega t$ in
collinear-beam scenario contributes only in the free motion of the
atomic beam along the $z$-axis and does not contain the radial
coordinate $r$ in contrast to the laser phase
$\zeta^{\scriptscriptstyle (\bot)}$ in crossed-beam scenario where
both $r$ and the azimuthal angle $\varphi$ are involved [cf.\
Eqs.~(\ref{lin-pol-col})-(\ref{lin-pol-cros})]. It is the presence
of $r$ and $\varphi$ that enables one to explore an intriguing
dynamics of atomic twisted states in crossed-beam scenario as we
exhibit in details in our forthcoming discussion.

\subsubsection{Crossed-beam scenario}
\label{sec:cross-beam}

So far, we have investigated the twisted states of free
(\ref{free-twisted-state}) and of laser-driven atoms in the
collinear-beam scenario (\ref{beam-col}). These two states involve
the same, \textit{stationary} Bessel-dependency $J_{\ell} (\xi)$ in
the profile of the beam. Whereas, the crossed-beam scenario gives
rise to a time-dependent atomic beam profile. To investigate this we
shall evaluate the integral (\ref{Integral}) by employing the
solution (\ref{solution2}). This integration cannot be carried out
by means of exact methods since the reduced quantities
$\alpha^{\scriptscriptstyle (\bot)}$, $\beta^{\scriptscriptstyle
(\bot)}$ and $\Omega^{\scriptscriptstyle (\bot)}$ depend on the
$\phi$ angle [compare Eqs.~ (\ref{reduced-quant-col-beam}) and
(\ref{reduced-quant-cros-beam})]. However, if we consider a
\textit{non-relativistic regime} for the propagation of atomic beam
we can easily proceed analytically. For this purpose, we again
represent the scalar product in the form $ \bm p \cdot \bm r \,=\,
p_\bot r \cos \left( \phi-\varphi \right) + p_{||} z$ and evaluate
the integral (\ref{Integral}) first  with respect to transverse
$p_\bot$ and longitudinal $p_{||}$ components of the (linear)
momentum
\begin{eqnarray}
\nonumber
   \Psi_{\ell}^{\scriptscriptstyle (\bot)}  & = &
   \frac{e^{\frac{i}{\hbar} (p_{|| 0} z - {\cal E}_0 t )}}{4 \pi i^\ell}
   \\
   \label{exact-int}
   & \times & \int_0^{2 \pi}
   \!\!\!\!\!\! \mathrm{d}  \phi \, e^{i \ell \phi} e^{i \xi \cos \left( \phi - \varphi \right)}
   \! \sum_{n=1}^4 e^{i {\cal A}_n \zeta^{\scriptscriptstyle (\bot)}} \! \ketm{{\cal B}_n} , \quad
\end{eqnarray}
where the state vectors
\begin{eqnarray*}
    \ketm{{\cal B}_{1,2}} \equiv \ketm{a} \, , \;
         \ketm{{\cal B}_{3,4}} \equiv \pm  e^{i\phi_{d_z}} \ketm{b}
\end{eqnarray*}
refer to the upper and lower levels of the atom. For the sake of
brevity, moreover, we introduce the dimensionless parameters
\begin{eqnarray}
\label{A}
    {\cal A}_{1,2} \equiv - \alpha_{0}^{\scriptscriptstyle (\bot)} \pm
    \Omega_{0}^{\scriptscriptstyle (\bot)} / 2
    \, , \;
    {\cal A}_{3,4} \equiv - \beta_{0}^{\scriptscriptstyle (\bot)} \pm
    \Omega_{0}^{\scriptscriptstyle (\bot)} / 2 \, ,
\end{eqnarray}
which carry information about the internal structure of the atom and
the strength of the atom-field coupling. These parameters are given
in terms of the reduced quantities $\alpha_{0}^{\scriptscriptstyle
(\bot)}$, $\beta_{0}^{\scriptscriptstyle (\bot)}$ and
$\Omega_{0}^{\scriptscriptstyle (\bot)}$ and are taken at the cone
surface $p~=~p_0$.

Now we apply our assumption about the non-relativistic regime for
the propagation of atomic beam and Taylor expand ${\cal A}_n$ ($n =
1 .. 4$) by keeping only the terms up to the first power in $v_{\bot
0} / c$
\begin{eqnarray*}
    {\cal A}_n \approx
    {\cal C}_n \left( 1 + \frac{v_{\bot 0}}{c} \cos (\phi_L - \phi) \right) \, ,
\end{eqnarray*}
where
\begin{eqnarray*}
    {\cal C}_{1,2} \equiv (E_a \mp \hbar \Omega_{R_z}/2)/(\hbar \omega) \, , \,
    {\cal C}_{3,4} \equiv (E_b \mp \hbar \Omega_{R_z}/2)/(\hbar \omega)
\end{eqnarray*}
are dimensionless constant energies normalized to the photon energy.
This expansion enables one to re-write the wave function
(\ref{exact-int}) in a desired form
\begin{eqnarray}
\nonumber \Psi_{\ell}^{\scriptscriptstyle (\bot)}  & = &
   \frac{e^{\frac{i}{\hbar} (p_{|| 0} z - {\cal E}_0 t )}}{4 \pi i^\ell}
     \\
\label{intermediate-int}
           & \times &
           \sum_{n=1}^4 e^{i {\cal C}_n \zeta^{\scriptscriptstyle (\bot)}}
      \int_0^{2 \pi}   \!\!\!\!\!\! \mathrm{d}
           \phi \, e^{i \ell \phi} e^{i {\cal X}_n \!  \cos \phi} e^{i {\cal Y}_n \! \sin \phi}
   \ketm{{\cal B}_n} \quad
\end{eqnarray}
which is appropriate for analytical integration. Here the
dimensionless \textit{coordinates}
\begin{eqnarray}
\nonumber
    {\cal X}_n  \left( \xi, \varphi, t \right) & \equiv &
    \xi \cos \varphi + {\cal C}_n \zeta^{\scriptscriptstyle (\bot)} \frac{v_{\bot 0}}{c} \cos \phi_L \, ,
    \\
\label{P,Q}
    {\cal Y}_n  \left( \xi, \varphi, t \right) & \equiv &
    \xi \sin \varphi + {\cal C}_n \zeta^{\scriptscriptstyle (\bot)} \frac{v_{\bot 0}}{c} \sin \phi_L
\end{eqnarray}
contain the (time-dependent) laser phase
\begin{eqnarray}
\label{laser-phase-cros}
    \zeta^{\scriptscriptstyle (\bot)}
    =  \frac{\hbar k}{p_{\bot 0}} \xi \cos \left( \phi_L - \varphi \right)  - \omega t \, , \quad k \equiv k_{\bot}
\end{eqnarray}
that is independent of the integration variable $\phi$.

In order to calculate the integral (\ref{intermediate-int}) we
introduce a new system of `cartesian' $\bm R_n \equiv \left({\cal
X}_n, {\cal Y}_n, {\cal Z} \right)$ and `cylindrical' $\bm R_n
\equiv \left(\Xi_n, \Phi_n, {\cal Z} \right)$ coordinates with
standard transformations
\begin{eqnarray}
\label{new-coord}
    {\cal X}_n = \Xi_n \cos \Phi_n \, , \quad
    {\cal Y}_n = \Xi_n \sin \Phi_n \, , \quad
    {\cal Z} = p_{\bot 0} z/\hbar \, , \quad\quad
\end{eqnarray}
where $\Xi_n  =  \sqrt{{\cal X}_n^2 + {\cal Y}_n^2}$ is the `radial'
coordinate and $\Phi_n$ is the `azimuthal angle' in the $( {\cal
X}_n , {\cal Y}_n)$-plane. By inserting these new coordinates in Eq.
(\ref{intermediate-int}) and making use of a simple trigonometric
relation, we can exploit the integral representation of the Bessel
function
\begin{eqnarray*}
   \int_0^{2 \pi} \!\!\!\!\!\! \mathrm{d}  \phi \,
        e^{i \ell \phi} e^{i \Xi_n \cos ( \phi - \Phi_n )}
   = 2 \pi i^\ell e^{i \ell \Phi_n} J_\ell ( \Xi_n )
\end{eqnarray*}
and obtain the final form of the twisted state of laser-driven
two-level atoms in crossed-beam scenario
\begin{eqnarray}
\label{beam-cros}
   \Psi_{\ell}^{\scriptscriptstyle (\bot)} =
   \frac{1}{2} e^{\frac{i}{\hbar} (p_{|| 0} z - {\cal E}_0 t )}
   \sum_{n=1}^4 e^{i {\cal C}_n \zeta^{\scriptscriptstyle (\bot)}}
     e^{i \ell \Phi_n} J_\ell ( \Xi_n )
   \ketm{{\cal B}_n}  \, . \quad\quad
\end{eqnarray}
It is important to note that we recover the free twisted state
(\ref{free-twisted-state}) if we switch off the laser field (i.e.
$\zeta^{\scriptscriptstyle (\bot)} \rightarrow 0$), since in this
case the coordinates $\Xi_n$ and $\Phi_n$ coincide with $\xi$ and
$\varphi$, respectively, as $ ( {\cal X}_n , {\cal Y}_n )
\rightarrow ( p_{\bot 0} x / \hbar, p_{\bot 0} y / \hbar)$ [cf.\
Eq.~(\ref{P,Q})].

Since no restriction has been made on the longitudinal $p_{|| 0}$
and transverse components $p_{\bot 0}$ of the atom momentum, the
states (\ref{beam-col}) and (\ref{beam-cros}) apply generally for
scalar Bessel beams of two-level atoms \textit{beyond} the typical
paraxial approximation for both, the collinear- and crossed-beam
scenarios. As we stressed above, we are not interested in the
normalization of these states since it does not provide too much
insight in the overall dynamics of the beam. To give a hint,
however, we would like to point out that these beams, can be
normalized, on one hand, by integrating them in a large, but a
finite cylindrical volume, as already mentioned above and also done
in \cite{Jentschura:11} for twisted photons. On the other hand, we
could regularize the Dirac $\delta$-function, which arises after the
integration of squared Bessel functions over the whole space, by
rigorously re-defining it as a limit of regular functions (e.g.
Gaussian), as also mentioned in \cite{Bliokh:11} for electron Bessel
beams.

In the last discussion of this section, we put the emphasis on
finding an integral of ``motion" that describes the propagation of
the laser-driven twisted atom in the crossed-beam scenario. A
similar operator description of angular momentum properties of light
Bessel beams has been done in \cite{Bliokh:10}. Given that the
photon energy and momentum are much less than the atom rest energy
and transverse momentum, respectively, i.e. $ \hbar \omega / (m c^2)
\ll 1$ and $\hbar k / (m v_{\bot 0}) \ll 1$, we replace (\ref{P,Q})
with the following approximate relations
\begin{eqnarray}
\nonumber
    {\cal X}_n & \approx &
    \xi \cos \varphi - {\cal C}_n v_{\bot 0} k t \cos \phi_L \, ,
    \\
\label{P,Q-approx}
    {\cal Y}_n  & \approx &
    \xi \sin \varphi - {\cal C}_n v_{\bot 0} k t \sin \phi_L \, ,
\end{eqnarray}
that are valid with high accuracy for a wide range of frequencies,
from infrared to ultraviolet regions. The wave function
(\ref{beam-cros}), therefore, can be re-written as
\begin{eqnarray}
\label{beam-cros-approx}
   \Psi_{\ell}^{\scriptscriptstyle (\bot)} & \approx & \sum_{n=1}^4
   \Psi_{\ell}^{\scriptscriptstyle (n)} \, ,
\\
\label{beam-component}
   \Psi_{\ell}^{\scriptscriptstyle (n)} & \equiv &
   \frac{1}{2} e^{\frac{i}{\hbar} (p_{|| 0} z - ({\cal E}_0 + {\cal C}_n \hbar \omega ) t )}
   e^{i \ell \Phi_n} J_\ell ( \Xi_n )
   \ketm{{\cal B}_n} \, , \quad
\end{eqnarray}
where we have replaced the laser phase (\ref{laser-phase-cros}) with
$\zeta^{\scriptscriptstyle (\bot)} \approx - \omega t$. One can
immediately recognize that the exponential $e^{i p_{|| 0} z/ \hbar}$
describes the diffraction-free propagation of the beam along the
$z$-axis. Moreover, this free propagation occurs as a sum of four
scalar Bessel modes $\Psi_{\ell}^{\scriptscriptstyle (n)}$, each of
which carries a quasi-energy ${\cal E}_0 + {\cal C}_n \hbar \omega$
because the atom is dressed by the field.

The notable difference between the twisted states of laser-driven
(also of free) atoms in collinear- and crossed-beam scenarios is
that here the Bessel function depends on the new coordinate $\Xi_n$
and, consequently, on time (compare Eqs.~(\ref{beam-col}),
(\ref{free-twisted-state}) with Eq.~(\ref{beam-cros-approx})). Due
to this (different) dependency, the state (\ref{beam-cros-approx})
is no longer the eigenstate of the \textit{conventional} OAM
operator $\hat{\ell}_z = - i \hbar
\partial / \partial \varphi$, instead each of the modes (\ref{beam-component})
represents an eigenstate of the operator
\begin{eqnarray}
\label{L-operator}
    \hat{{\cal L}}_z^{{\scriptscriptstyle (n)}} \equiv
    - i \hbar \frac{\partial }{\partial \Phi_n} \, ,
\end{eqnarray}
an \textit{OAM} operator which acts in $( {\cal X}_n , {\cal Y}_n,
{\cal Z})$ configuration space, upon which our physical system
depends. It is easy to verify that this operator has the eigenvalue
$\hbar \ell$, i.e. $\hat{{\cal L}}_z^{{\scriptscriptstyle (n)}}
\Psi_{\ell}^{\scriptscriptstyle (n)} = \hbar \ell
\Psi_{\ell}^{\scriptscriptstyle (n)}$. Whereas, as mentioned above,
the states (\ref{free-twisted-state}) and (\ref{beam-col}) are the
eigenstates of $\hat{\ell}_z$ with the same eigenvalue $\hbar \ell$,
i.e. $\hat{\ell}_z \psi_\ell = \hbar \ell \psi_\ell $ and
$\hat{\ell}_z \Psi_{\ell}^{\scriptscriptstyle (||)} = \hbar \ell
\Psi_{\ell}^{\scriptscriptstyle (||)}$, respectively. In addition,
as one may expect the operator (\ref{L-operator}) coincides with the
operator $\hat{\ell}_z$ when the field is switched off, $\hat{{\cal
L}}^{\scriptscriptstyle (n)}_z \rightarrow \hat{\ell}_z$, (see also
Eq.\ (\ref{OAMin-new-coord})).

To get a deeper insight let us now calculate the mean value of the
conventional OAM operator averaged over the modes
$\Psi_{\ell}^{\scriptscriptstyle (n)}$. For this purpose, we express
$\hat{\ell}_z$ in terms of coordinates (\ref{new-coord}) by
employing the relations (\ref{P,Q-approx})
\begin{eqnarray}
\nonumber
    \hat{\ell}_z & = & - i \hbar \frac{\partial}{\partial \varphi} =
                                    - i \hbar
    \left( \frac{\partial \Phi_n}{\partial \varphi} \frac{\partial}{\partial \Phi_n}
    + \frac{\partial \Xi_n}{\partial \varphi} \frac{\partial}{\partial \Xi_n}
       \right)
\\
\nonumber
     &  = & - i \hbar  \left[
            \frac{\partial}{\partial \Phi_n} +
                    {\cal C}_n v_{\bot 0} k t
                      \frac{ \cos  \left( \Phi_n  -  \phi_L \right)}{\Xi_n}
                       \frac{\partial}{\partial \Phi_n} \right.
\\
\label{OAMin-new-coord}
     & + & \left.
                 {\cal C}_n v_{\bot 0} k t
                   \sin \! \left( \Phi_n  -  \phi_L \right)
                     \frac{\partial}{\partial \Xi_n}
                         \right] \, .
\end{eqnarray}
Let us note that a similar operator has been derived in
\cite{Karlovets:12} to describe the electron with non-zero OAM in
the presence of a strong laser field. Furthermore, for the (quantum
mechanical) mean value of $\hat{\ell}_z$, when normalized to the
overall `energy' of the mode (\ref{beam-component}), we obtain
\begin{eqnarray}
\label{mean-value}
     \langle \hat{\ell}_z \rangle =
     \frac{\displaystyle \int \!\!\! d \bm R_n
     \left( \Psi_{\ell}^{\scriptscriptstyle (n)} \right)^*
     \left( \hat{\ell}_z \Psi_{\ell}^{\scriptscriptstyle (n)} \right) }
     {\displaystyle \int \!\!\! d \bm R_n \left| \Psi_{\ell}^{\scriptscriptstyle (n)} \right|^2 }
                                     = \hbar \ell \, ,
\end{eqnarray}
where $d \bm R_n = \Xi_n d \Xi_n d \Phi_n d {\cal Z} $ is the
elementary cylindrical volume in configuration space which is
related to the elementary volume in position space via $d \bm r =
(\hbar / p_{\bot 0})^3 d \bm R_n$. The last two terms of the
operator (\ref{OAMin-new-coord}) do not contribute to the integral
after the integration over $\Phi_n$. As seen, the mean values of
both OAM operators coincide, $\langle \hat{\cal
L}^{\scriptscriptstyle (n)}_z \rangle = \langle \hat{\ell}_z
\rangle$. This means that, apart from the diffraction-free
propagation along $z$-axis, the state (\ref{beam-cros-approx})
describes a superposition of four modes (\ref{beam-component}) each
of which carries a non-zero OAM with respect to the same axis.
Moreover, the time-dependent profile of the beam reveals a
non-trivial dependency on the transverse coordinate $\xi$ and the
azimuthal angle $\varphi$. In the next subsection, we examine this
non-triviality in details and spatiotemporally characterize the
Bessel beams of two-level hydrogen and selected alkali-metal atoms
that are resonantly driven by the laser field without the level
damping.

\section{Spatial and temporal characterization of atomic Bessel beams}
\label{sec:Prob.dens}

So far we have built the twisted states (\ref{beam-col}) and
(\ref{beam-cros}), (\ref{beam-cros-approx}) of laser-driven
two-level atoms for collinear- and crossed-beam scenarios. These
states can be used in order to study the space- and time-dependent
profile of atomic beams. In this section, therefore, we put the
emphasis on the crossed-beam scenario and investigate how the radial
distribution and the time evolution of the probability density are
affected by the atomic beam velocity and the nuclear charge $Z$ when
the laser field is resonant to the atomic transition energy.

To proceed with our further discussion, we define the probability
density of atoms in a twisted state with a non-zero OAM $\hbar \ell$
as
\begin{eqnarray}
\label{density}
     \rho_{\ell} & = & \left| \Psi_\ell \right|^2 \, ,
\end{eqnarray}
such that $\rho = 1$ for beams (\ref{solution1}) and
(\ref{solution2}) of \textit{non-twisted} atoms. Depending on which
of the two scenarios occurs, the probability density (\ref{density})
acquires different forms. To show this, we substitute the wave
function (\ref{beam-col}) into Eq.~(\ref{density}) and obtain the
probability density for the collinear-beam scenario
\begin{eqnarray}
\label{density-col}
   \rho_{\ell}^{\scriptscriptstyle (||)}  =  J_\ell^2 \left( \xi \right) \, ,
\end{eqnarray}
as a function of only the dimensionless transverse coordinate $\xi$.
Figure \ref{fig.2} shows the `non-diffracting' distribution of this
probability density that coincides with the beam profile of free
twisted atoms (even if the laser field is still switched on). This
coincidence is due to the $\xi$-independent laser phase
(\ref{lin-pol-col}) as also pointed out in Subsection
\ref{sec:coll-beam}

Let us now calculate the probability density in crossed-beam
scenario. For this purpose, by inserting the state (\ref{beam-cros})
into Eq.~(\ref{density}), after straightforward derivations, we
obtain
\begin{eqnarray}
\label{density-cros}
    \rho_{\ell}^{\scriptscriptstyle (\bot)} = \varrho_{\ell} + \Delta_{\ell} \, ,
\end{eqnarray}
where the (space- and time-dependent) term
\begin{eqnarray}
\label{density-crosB}
     \varrho_{\ell} = \frac{1}{4} \sum_{n = 1}^4 J_\ell^2
                           \left( \Xi_n \right)
\end{eqnarray}
is responsible for the Bessel-squared-type distribution of the
probability density, and the term
\begin{eqnarray}
\nonumber
   \Delta_{\ell}
   & = &  \frac{1}{2}  \cos \! \left( \!  \frac{\Omega_{R_z}}{\omega} \zeta_\bot
              + \ell ( \Phi_2 - \Phi_1) \right)
              J_\ell \left( \Xi_1\right)
              J_\ell \left( \Xi_2 \right)
\\
\label{density-crosO}
   & - &  \frac{1}{2}  \cos \! \left( \!  \frac{\Omega_{R_z}}{\omega} \zeta_\bot
             + \ell ( \Phi_4 - \Phi_3) \right)
             J_\ell \left( \Xi_3 \right)
             J_\ell \left( \Xi_4 \right)  ,\quad
\end{eqnarray}
is a small summand that can be neglected with high accuracy under
properly tuned parameters of the ``atom + laser" system, as shown
below.

\begin{figure}
\includegraphics[width=0.48\textwidth]{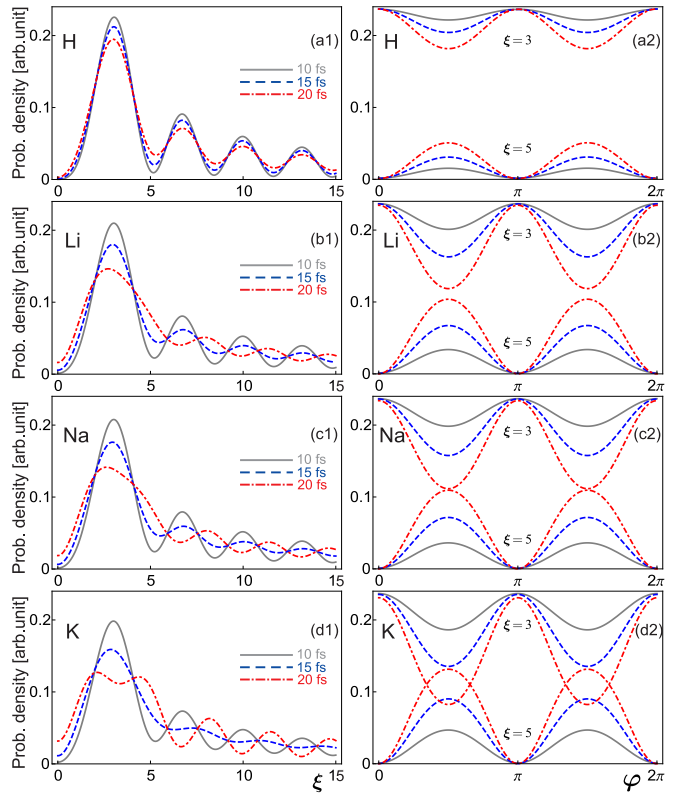}
\caption{(Color online). Distribution of probability density $
\rho_{\ell}^{\scriptscriptstyle (\bot)} $ (in arbitrary units) as a
function of dimensionless transverse coordinate $\xi = p_{\bot 0} r
/ \hbar$ (left panel) and azimuthal angle $\varphi$ (right panel)
for $\ell = 2$. The probability densities are shown for Bessel beams
of H (a1-2) with atomic velocity $1.4 \cdot 10^6$ cm/s and of Li
(b1-2), Na (c1-2), K (d1-2) with atomic velocity
 $0.7 \cdot 10^6$ cm/s for different propagation times $10$ fs
(gray solid curves), $15$ fs (blue dashed curves) and $20$ fs (red
dot-dashed curves).} \label{fig.3}
\end{figure}

In order to explore and exhibit the temporal and spatial
characteristics of atomic Bessel beams let us consider, for example,
two-level hydrogen, lithium, sodium and potassium, and assume that
these atoms are driven on the $1s \leftrightarrow 2p$, $2s
\leftrightarrow 2p$, $3s \leftrightarrow 3p$ and $4s \leftrightarrow
4p$ atomic transitions, respectively. For the sake of simplicity,
however, we here also suppose that \textit{no} decay occurs for
upper levels and, thus, that no damping applies in the
time-evolution of the probability amplitudes
(\ref{sol.formA.res})-(\ref{sol.formB.res}). For the $2p
\leftrightarrow 1s$ transition, the laser and Rabi frequencies for
hydrogen can be easily expressed as
\begin{eqnarray}
\label{2p1s}
    \omega & = & \frac{3 e^2}{8 a_0 \hbar}
    \, , \quad
    \Omega_{R_x} = \frac{2^7 a_0 e \varepsilon_0 }{3^5 \hbar}
    \, , \quad
    \Omega_{R_z} = \sqrt{2} \Omega_{R_x} \quad
\end{eqnarray}
by using the well-known properties of hydrogen-like ions
\cite{Bransden:83}. For alkali-metal atoms, we make use of the known
values of their spectrum \cite{Parsons:67} and dipole matrix
elements \cite{Metcalf:99}. Moreover, we here restrict our
discussion to low- and medium-$Z$ atoms since the interaction
Hamiltonian (\ref{int.Ham1})-(\ref{int.Ham2}) is valid only within
the LWA, i.e. when the radiation wavelength exceeds the atomic
sizes.

\begin{figure}[ht]
\includegraphics[width=0.49\textwidth]{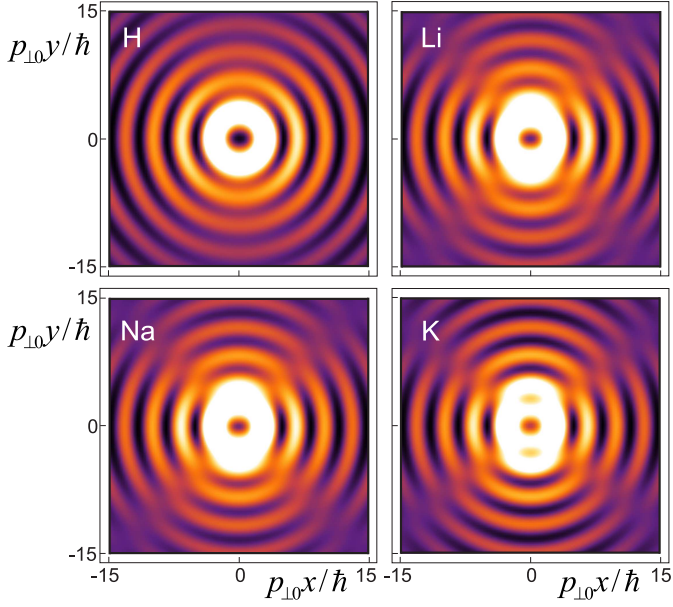} \caption{(Color
online). `Snapshot' of atomic Bessel beam profiles at $t = 20$ fs
after the laser, which drives the atom, is activated. The laser wave
propagates along the positive direction of $y$-axis, the
polarization vector points toward the observer. The parameters of
the ``atom + laser" system and the variation of colors are the same
as in Figures \ref{fig.3} and \ref{fig.2}, respectively.}
\label{fig.4}
\end{figure}

After we have specified the type of a laser-driven two-level atom we
are ready to explore both, the spatial and temporal characteristics
of atomic Bessel beams. When the laser field is switched off, i.e.
for $t = 0$, both curves for collinear- and crossed-beam scenarios
coincide [cf.\ Figure~\ref{fig.2}]. This is quite expected since in
the absence of the laser radiation only a free Bessel beam
propagates, as one might also expect due to the initial conditions
(\ref{cond-for-lin}). Once the laser is switched on, the atomic beam
starts evolving in the field and changing its conventional
Bessel-squared-shape. Figure \ref{fig.3} displays the probability
density profile (i) at various dimensionless transverse coordinates
ranging from $0$ to $15$ at given azimuthal angle $\varphi = \pi /
3$ (left panel) as well as (ii) at various azimuthal angles ranging
from $0$ to $2 \pi$ at given (two different values of) transverse
coordinate $\xi = 3$ and $\xi = 5$ (right panel). In particular,
results are shown for the nonparaxial atomic beam with transverse
momentum $p_{\bot 0} = p_{0} / 5$ and OAM $\hbar \ell = 2 \hbar$ as
well as for the field strength $\varepsilon = 4 $ GV/cm and laser
propagation angle $\phi_L = \pi / 2$ for three different evolution
times, $10$, $15$ and $20$ fs. Moreover, Figure \ref{fig.4} combines
both the $\xi$- and $\varphi$-dependencies and shows the
\textit{actual} profile of Bessel beams of the same atoms when they
propagate $20$ fs in the field.

As seen in Figure \ref{fig.3}, the deviation of curves from each
other increases the longer the atom propagates in the laser field.
This deviation is caused by all four Bessel modes in
Eq.~(\ref{density-crosB}) containing the arguments
\begin{eqnarray}
\nonumber
    \Xi_n & = & \sqrt{\xi^2 + 2 \xi {\cal C}_n \zeta^{\scriptscriptstyle (\bot)} \frac{v_{\bot 0}}{c}
                             \cos \left( \phi_L - \varphi \right) +
                         \left( {\cal C}_n \zeta^{\scriptscriptstyle (\bot)} \frac{v_{\bot 0}}{c}  \right)^2}
\\
\label{argument}
    & \approx &    \sqrt{\xi^2 - 2 \xi {\cal C}_n v_{\bot 0}k t
                         \cos \left( \phi_L - \varphi \right)}  \, .
\end{eqnarray}
The time factor $2 \xi {\cal C}_n v_{\bot 0} k t \cos \left( \phi_L
- \varphi \right)$, which involves both the transition energy and
the atom-laser coupling strength, can lead to an
\textit{enhancement} of the second maximum of the beam profile
\begin{figure}[ht]
\includegraphics[width=0.49\textwidth]{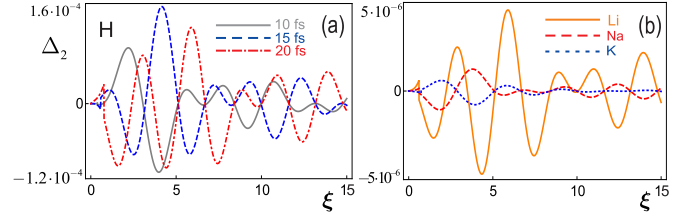} \caption{(Color
online). Distribution of $\Delta_\ell$ for $\ell = 2$ as a function
of dimensionless transverse coordinate $\xi = p_{\bot 0} r / \hbar$
for hydrogen with different propagation times (left panel) and for
alkali-metal atoms with an evolution time $10$ fs (right panel).}
\label{fig.5}
\end{figure}
during its evolution in the laser field. This enhancement depends
crucially on how the parameters of the ``atom + laser" system are
tuned. For instance, when the beam of potassium is evolved in the
field within $20$ fs, the first two maxima become of the same order
[cf.\ Figure \ref{fig.3} (d1)]. This can be clearly seen also (i)
from Figure \ref{fig.3} (d2) in the region where the red dot-dashed
curves intersect for $\xi = 3$ and $\xi = 5$ and (ii) from Figure
\ref{fig.4} especially for the azimuthal angle $\varphi \sim \pi/2$
for which the first two maxima are well separated (white areas for
the potassium profile in the vertical direction). In addition, the
probability density (\ref{density-crosB}) for exact $\varphi = \pi /
2$ is discussed in \cite{Hayrapetyan:13}.

Figure \ref{fig.3} exhibits also another intriguing feature of the
beam profile in the crossed-beam scenario (which, in fact,
eventually leads to the enhancement of the second maximum). Since
the term ${\cal C}_n v_{\bot 0} k t $ reveals different values for
different atoms, this $Z$-dependency gives rise to a
\textit{field-induced `spread'} of the Bessel-squared-shape of the
atomic probability density [cf.\ Figure \ref{fig.3} (b1)-(d1)]. This
behavior is rather universal and remains the same for all $z$, i.e.
along the atom propagation axis. Indeed, as the nuclear charge
increases the white areas in Figure \ref{fig.4} start to
symmetrically spread along and against the propagation direction of
the field, meanwhile keeping their shape constant along the
$z$-axis. All these non-trivial spatiotemporal characteristics are
caused by the coherent interaction of atomic and laser beams and are
quantitatively reflected both in the $Z$-dependent atomic transition
energy and the atom-field interaction strength, i.e. the Rabi
frequency. In addition, we would like to note that the term
$\Delta_\ell$ does not contribute in the radial distribution of the
probability density because it is effectively zero for an evolution
time varying in $0, .. 100$ fs and for low- and medium-$Z$ atoms
[cf.\ Figure~\ref{fig.5}].

As, for example, in the case of optical \cite{Kuntz:09,Bliokh:12}
and electron \cite{Bliokh:12} Bessel beams, we have spatially and
temporally characterized the atomic Bessel beams that are driven by
the laser field. In contrast to these studies, radiation-assisted
atomic beams, that carry non-zero OAM, acquire a special type of
behavior, as illustrated in Figures \ref{fig.3}-\ref{fig.5}, due to
the coherent coupling of an atom to the laser field. To study these
properties of the laser-driven Bessel beams experimentally, we hope
that similar methods as for the creation of electron vortex beams
via \textit{nanofabricated fork-like hologram} [cf.\
Refs.~\cite{Verbeeck:10-11,McMorran:11}] will be be developed for
the atomic systems. Moreover, for neutral atoms, that barely
interact with the matter, we believe that this can be achieved by
means of (i) atomic microscopes which deliver resolution of the
order of few nm and (ii) laser systems which provide a resolution of
the order of $10$ fs.

\section{\label{Conclusion}Summary and conclusions}

The twisted states of laser-driven two-level atoms have been built
and investigated, especially, in crossed-beam scenario when the
laser and atomic beams are perpendicular to each other. In more
detail, the interaction of a two-level atom with a linearly
polarized electromagnetic field has been described by using the
space- and time-dependent laser phase for solving the Schr\"odinger
equation in a similar way as known from relativistic quantum theory
of electron. Exact analytical solution to the Schr\"odinger equation
was found within the eikonal, rotating-wave and long-wave
approximations (to deal with fields nearly resonant to the two-level
excitation energy and with wavelengths larger than the atomic size).
Our treatment enables one to construct a twisted state of
laser-driven two-level atoms with their well defined energy,
transverse and longitudinal momentum components as well as the
projection of the orbital angular momentum along the propagation
direction. By making use of these states, detailed calculations have
been performed for the distribution of the probability density of
hydrogen, lithium, sodium and potassium for the $1s \leftrightarrow
2p$, $2s \leftrightarrow 2p$, $3s \leftrightarrow 3p$ and $4s
\leftrightarrow 4p$ atomic transitions, respectively,
\textit{without} (level) damping. For the crossed-beam scenario, we
have exhibited a non-trivial, Bessel-squared-type behavior of the
beam profile that applies for both, paraxial and nonparaxial regimes
and depends on time. We have also shown that a possible enhancement
of the second maximum of probability density may occur under a
specific choice of laser and atom parameters, such as the atom
evolution time, nuclear charge, atomic velocity, the laser frequency
and the electric field strength.

Though emphasis was placed on the interaction of two-level atoms
with external (monochromatic) fields, the theory in this work is
applicable also for three- and multi-level atoms of different
configurations, including $\Lambda -$, $V -$ and $\Sigma -$ type
atoms. Moreover, this theory can be extended quite easily to a
(general) elliptically polarized monochromatic field as well as to a
\textit{standing} wave $\bm{E} \,=\, (\varepsilon
\cos(\bm{k}\bm{r}-\omega t)+\varepsilon \cos(\bm{k}\bm{r}+\omega
t),0,0)$. The latter case will enable one, for instance, to deal
also with \textit{elastic} interactions between the atom and the
field. In the future, moreover, it seems desirable to us (and
possible) to construct a unitary operator of the form
\begin{eqnarray*}
   U_{\ell}^{\scriptscriptstyle (||)} \! \left(t,t^\prime\right) & = &
   \frac{\cos ( \Omega_{0}^{\scriptscriptstyle (||)} \zeta^{\scriptscriptstyle (||)} / 2 )}
   {\cos ( \Omega_{0}^{\scriptscriptstyle (||)} \zeta^{\prime \scriptscriptstyle (||)} / 2 )} \;
   e^{i\alpha_{0}^{\scriptscriptstyle (||)} \omega\left(t-t^\prime\right)} \:\ketm{a}\bram{a}
\\
   & + &
   \frac{\sin ( \Omega_{0}^{\scriptscriptstyle (||)} \zeta^{\scriptscriptstyle (||)} / 2 )}
   {\sin ( \Omega_{0}^{\scriptscriptstyle (||)} \zeta^{\prime \scriptscriptstyle (||)} / 2 )} \;
   e^{i\beta_{0}^{\scriptscriptstyle (||)} \omega\left(t-t^\prime\right)}  \:\ketm{b}\bram{b} \,
\end{eqnarray*}
with the phase $\zeta^{\prime \scriptscriptstyle (||)} \equiv k_{||}
z - \omega t^\prime$ at some initial state in the collinear-beam
scenario, in order to describe collision and scattering phenomena
with atomic Bessel beams. Such evolution operators would be useful
also for investigation of form factors of Bessel beams of
radiation-assisted two-level systems, such as molecules, atoms or
even nuclei~\cite{Povh:08}.

\begin{acknowledgement}

A.G.H.\ thanks Dr. Marco Ornigotti and Dr. Filippo Fratini for
useful comments and acknowledges the support from the GSI
Helmholtzzentrum and the University of Heidelberg. O.M. and A.S.
acknowledge support from the Helmholtz Gemeinschaft and GSI
(Nachwuchsgruppe VH-NG-421).

\end{acknowledgement}

\appendix
\numberwithin{equation}{section}
\addcontentsline{toc}{section}{Appendix~\ref{app:derivation}}
\section{Derivation of solutions of the Schr\"odinger equation} \label{app:derivation}

To determine the probability amplitudes $\psi_a$ and $\psi_b$ for
the atomic states \ket{a} and \ket{b}, we take into account that the
overall dynamics of the atom-laser system remains the same for the
collinear- and crossed-beam scenarios  (compare Eqs.
(\ref{int.Ham1})-(\ref{int.Ham2})), though they imply different
coupling strengths and phases in the interaction Hamiltonian. For
these similar Hamiltonians, in our further calculations we will
adopt generic notations $\Omega_R$, $\phi_d$ and $\zeta$ in order to
replace the Rabi frequencies $\Omega_{R_x}$, $\Omega_{R_z}$, the
dipole matrix element exponentials $\phi_{d_x}$, $\phi_{d_z}$  and
the laser phases $\zeta_{||}$, $\zeta_\bot$, respectively. The
replacement of these laser phases is justified since the solution
which we construct does not depend on any preferred direction of the
laser wave vector.

In order to describe the dynamics of laser-driven two-level atoms we
take into account that the laser phase $\zeta$ contains both, the
time- and space-variables and, therefore, enables one to express the
corresponding partial deri-vatives by the total derivative with
regard to this phase. Based on this mathematical trick we can
re-write the Schr\"o-dinger equation as an ordinary differential
equation in a similar way as the Dirac equation has been examined
earlier by Volkov \cite{Berestetskii:82} and Skobelev
\cite{Skobelev:88} who found solutions for the (relativistic) motion
of electrons and neutrons, respectively. Thus, by making use of this
technique and substituting ansatz (\ref{psi}) into the
time-dependent Schr\"odinger equation (\ref{Sch.eq}), we obtain the
two coupled equations
\begin{small}
\begin{eqnarray*}
   \frac{\hbar^2 k^2}{2m}\psi_a'' +
   i\hbar \left( \bm{v} \cdot \bm{k} - \omega \right) \psi_a' -
    E_a \psi_a
  + \hbar\Omega_R \, e^{-i\phi_d} \cos\zeta \,\psi_b & = & 0
   \\[0.2cm]
   \frac{\hbar^2 k^2}{2m} \psi_b'' +
   i\hbar \left( \bm{v} \cdot \bm{k} - \omega \right) \psi_b' -
   + E_b \psi_b
  + \hbar\Omega_R \, e^{ i\phi_d} \cos\zeta \,\psi_a & = & 0 \, ,
\end{eqnarray*}
\end{small}
where the prime refers to a derivation with regard to the phase
$\zeta$ and $\bm{v} \:=\: \bm{p}/m$ denotes the center-of-mass
velocity of the atom with the non-relativistic energy ${\cal E}
\:=\: p^2 / (2m)$. Similar techniques have been applied more
recently also to the Mott scattering of an electron in the presence
of intense single-mode laser fields \cite{Szym:97}, both within the
relativistic and non-relativistic regime. For non-relativistic
electrons and neutrons, moreover, such an approach was taken in
Refs.~\cite{Mkrtchyan:09} and \cite{Mkrtchyan-Grigoryan},
respectively.

Owing to the (large) mass $m$ of the atom, whose rest energy $m c^2$
is much larger than the photon energy $\hbar \omega$, we typically
have $\hbar^2k^2/(2m\hbar\omega) \leq 10^{-10}$ even for ultraviolet
frequencies and can hence make use of the EA \cite{Landau:2}. In
this approximation, we will drop the first terms in the last system
of equations and rewrite them in the form
\begin{eqnarray}
\label{Schpsi_a}
   \psi_a' \,+\, i\alpha\psi_a & = &
   i\Omega \, e^{-i\phi_d}\: \cos\zeta \,\psi_b,
   \\[0.1cm]
\label{Schpsi_b}
   \psi_b' \,+\, i\beta\psi_b  & = &
   i\Omega \, e^{ i\phi_d}\: \cos\zeta \,\psi_a,
\end{eqnarray}
by introducing the following notations
\begin{eqnarray}
\label{reduced-quant}
   \alpha \equiv \frac{E_a}{\hbar\left(\bm{v}\bm{k}-\omega\right)} \, , \,
   \beta   \equiv  \frac{E_b}{\hbar\left(\bm{v}\bm{k}-\omega\right)} \, , \,
   \Omega \equiv \frac{\Omega_R}{\bm{v}\bm{k}-\omega} \, . \quad
\end{eqnarray}
The denominator $\bm{v} \cdot \bm{k}-\omega$ hereby illustrates the
Doppler shifted radiation frequency as seen by the moving atom
\cite{Landau:2}.

Eqs.~(\ref{Schpsi_a}) and (\ref{Schpsi_b}) describe the evolution of
the probability amplitudes in EA. To find solutions for these two
first-order equations, we may use the ansatz
\begin{eqnarray}
\label{psi_a}
   \psi_a & = & A (\zeta) \: e^{-i\alpha\zeta},
   \\
\label{psi_b}
   \psi_b & = & B (\zeta) \: e^{-i\beta\zeta}
\end{eqnarray}
to bring them into the simpler form
\begin{eqnarray}
\label{SchABex.1a}
   A' & = & i\Omega \, e^{-i\phi_d} \,\cos\zeta \,
            e^{i\left(\alpha-\beta\right)\zeta} \, B,
   \\
\label{SchABex.1b}
   B' & = & i\Omega \, e^{i\phi_d}  \,\cos\zeta \,
            e^{-i\left(\alpha-\beta\right)\zeta} \, A \, ,
\end{eqnarray}
in which the second terms on the left-hand side of Eqs.
(\ref{Schpsi_a}) and (\ref{Schpsi_b}) have been eliminated. As we
will see below, this re-definition (\ref{psi_a}) and (\ref{psi_b})
of the probability amplitudes facilitates the integration of
Eqs.~(\ref{SchABex.1a})-(\ref{SchABex.1b}). In addition, we could
also \textit{decouple} these two equations by taking the second
derivative with regard to the phase $\zeta$,
\begin{eqnarray}
\label{SchABex.2a}
  A'' + \left( -i (\alpha-\beta) + \tan \zeta \right) A' +
  \Omega^2\cos^2\zeta \, A & = & 0 \,
  \\[0.1cm]
\label{SchABex.2b}
  B'' + \left(  i (\alpha-\beta) +  \tan \zeta \right) B' +
  \Omega^2\cos^2\zeta \, B & = & 0 \, , \quad
\end{eqnarray}
and for which solutions are known in terms of hypergeometric
functions \cite{Gradshtein:00}. However, in this work we are
interested in laser frequencies which are nearly resonant to atomic
transition frequency. Therefore, we apply RWA to obtain solutions to
Eqs.~(\ref{SchABex.2a}) and (\ref{SchABex.2b}) in terms of
elementary functions.

In our further discussion, as usual, we consider frequencies (of the
radiation field) which are in resonance with or nearly resonant to
the atomic excitation, $\hbar \omega \,\approx\, E_a - E_b$. In such
a \textit{resonance} regime, it is justified to apply the RWA for
which the exact solutions can be found for Eqs.~(\ref{SchABex.1a})
and (\ref{SchABex.1b}). If the counter-rotating terms proportional
to $\exp\left[\pm i\left(\alpha-\beta-1\right)\zeta\right]$ are
ignored on the right-hand side, these equations take the form
\begin{eqnarray}
\label{SchAB}
   A' & = & \frac{i\Omega}{2} e^{-i\phi_d}
   e^{i\left(\alpha-\beta+1\right)\zeta} \, B \, ,
   \\[0.1cm]
   B' & = & \frac{i\Omega}{2} e^{i\phi_d}
   e^{-i\left(\alpha-\beta+1\right)\zeta} \, A \,.
\label{SchBA}
\end{eqnarray}
An (exact) solution for $A$ and $B$ is then given by
\begin{eqnarray}
\nonumber
    \left( \mu_1-\mu_2 \right) A \left( \zeta \right) & = &
   -\left[A (0) \mu_2 - \frac{i\Omega}{2} e^{-i\phi_d} B (0) \right]
   e^{\mu_1\zeta} \quad \quad
   \\[0.1cm]
\label{sol.formA}
    & + & \left[A (0) \mu_1 - \frac{i\Omega}{2} e^{-i\phi_d} B (0) \right]
   e^{\mu_2\zeta} \, ,
\end{eqnarray}
\begin{eqnarray}
\nonumber
    \left( \mu_1-\mu_2 \right) B \left( \zeta \right)& = & \hspace*{0.35cm}
    \left[B (0) \mu_1 + \frac{i\Omega}{2} e^{ i\phi_d} A (0) \right]
   e^{-\mu_2\zeta} \quad \quad
      \\[0.1cm]
\label{sol.formB}
   & - & \left[B (0) \mu_2 + \frac{i\Omega}{2} e^{ i\phi_d} A (0) \right]
   e^{-\mu_1\zeta} \, ,
\end{eqnarray}
with
\begin{eqnarray}
\label{mu_1,2}
   \mu_{1,2}=\frac{i}{2}\left(\alpha-\beta+1\pm \delta\right) \, ,
\\[0.1cm]
\label{delta}
   \delta\equiv \sqrt{\left(\alpha-\beta+1\right)^2+\Omega^2} \, ,
\end{eqnarray}
and where $A\left(0\right)$ and $B\left(0\right)$ refer to
``initial" conditions with regard to the laser phase, $\zeta = 0$.
These initial conditions are fulfilled, for instance, at the origin
$\bm{r}=0, \,t=0$ of the space and time coordinates. Owing to the
existence of the phase $\zeta$, however, the solutions
(\ref{sol.formA}) and (\ref{sol.formB}) are more general and can be
analyzed in order to explore the time- and space-dependency of the
probability amplitudes explicitly, if one wishes to take into
account the wave vector of the laser field and the motion of the
atom as a whole. For the atom at rest ($\bm{p} = 0$) and if we also
ignore the $\bm k$-dependence of the laser beam, i.e. simply perform
a replacement $\zeta \rightarrow -\omega t$, the solutions
(\ref{sol.formA}) and (\ref{sol.formB}) simplify to
\begin{eqnarray*}
   A \left( t \right) & = &
   \left\{ A (0) \left[\cos{\frac{\Lambda_2 t}{2}}
                       - i \frac{\Lambda_1}{\Lambda_2} \sin{\frac{\Lambda_2 t}{2}}
                \right] \right.
\\[0.1cm]
\nonumber
          & + & \left. i \frac{\Omega_R}{\Lambda_2} \, e^{ -i\phi_d} \,
              B (0) \sin{\frac{\Lambda_2 t}{2}}
   \right\} \; e^{ i\frac{\Lambda_1 t}{2}} \, ,
   \\[0.2cm]
   B \left( t \right) & = &
   \left\{ B (0) \left[ \cos{\frac{\Lambda_2 t}{2}}
                       + i \frac{\Lambda_1}{\Lambda_2} \sin{\frac{\Lambda_2 t}{2}}
                 \right] \right.
\\[0.1cm]
\nonumber
           & + & \left. i \frac{\Omega_R}{\Lambda_2} \, e^{ i\phi_d} \,
              B (0) \sin{\frac{\Lambda_2 t}{2}}
    \right\} \;  e^{ -i\frac{\Lambda_1 t}{2}}
\end{eqnarray*}
with $\hbar \Lambda_1 \,=\, E_a - E_b - \hbar \omega$ and
$\Lambda_2^2 \,=\, \Lambda_1^2+\Omega_R^2$, and are in full
agreement with standard texts on the two-level atom
\cite{Scully/Zubairy:01}. Using the Eqs.~(\ref{sol.formA}) and
(\ref{sol.formB}), moreover, the conservation of the overall
probability of the atom, namely of being in one of the two states,
$\left|\psi_e\right|^2+\left|\psi_g\right|^2 \:=\: 1$, can be
verified quite easily.

Although an oscillation of the probability amplitudes $\psi_a$ and
$\psi_b$ can be seen already from ansatz
(\ref{psi_a})-(\ref{psi_b}), further insights are obtained if we
specify the ``initial" conditions for these probability amplitudes
and modulate the field in a resonance regime with the atomic
transition frequencies. Therefore, two remarks are in order here
before we shall further examine the solutions
(\ref{sol.formA})-(\ref{sol.formB}) for the amplitudes $A(\zeta)$
and $B(\zeta)$. First, the initial condition for $\zeta = 0$ should
be chosen properly. If we assume the atom initially to be in the
upper state, we have
\begin{eqnarray}
\label{cond-for-lin}
    A\left(0\right) & = & 1 \, , \qquad  B\left(0\right) \;=\; 0 \, ,
\end{eqnarray}
for the interaction of the atom with linearly polarized field
\cite{Scully/Zubairy:01}.  Second, we still have some freedom in
general of how to choose the physical parameters, such as the
momentum $\bm{p} $ of the atom, its energies $E_a,\: E_b$ of the
upper and lower states, the frequency $\omega$ of the coupling field
as well as the Rabi frequency $\Omega_R$. Apart from a suitable
choice of the two-level atom, these parameters are often controlled
by the intensity and the propagation direction of the external
field(s) acting upon the atom. For example, we may readily fulfill
the \textit{resonance} condition by assuming a field with frequency
$\omega$, so that
\begin{eqnarray*}
   E_a - E_b
   & = & \hbar\omega_{\rm\, eff}
   \;\equiv\; \hbar \omega\, \left( 1+\frac{\bm{v} \cdot \bm{n}}{c}\right) \, ,
\end{eqnarray*}
and where $\bm n$ is the unit vector along the propagation direction
of the field as used before. This resonance condition gives rise to
the simple relation
\begin{eqnarray}
\label{res}
    \alpha-\beta+1 & = & 0
\end{eqnarray}
for the reduced quantities (\ref{reduced-quant}).

If we substitute the initial condition (\ref{cond-for-lin}) into the
solutions (\ref{sol.formA})-(\ref{sol.formB}) for a linearly
polarized field, the \textit{phase}-dependent probability amplitudes
read as
\begin{eqnarray*}
   A & = & \frac{i\mu_2}{\delta} \: e^{\mu_1\zeta} -
   \frac{i\mu_1}{\delta} \: e^{\mu_2\zeta} \, ,
   \\[0.1cm]
   B & = & \frac{\Omega}{2\delta} \: e^{ i\phi_d} e^{-\mu_2\zeta} -
   \frac{\Omega}{2\delta} e^{ i\phi_d} \: e^{-\mu_1\zeta} \, ,
\end{eqnarray*}
and, together with the resonance condition (\ref{res}), take the
simple form
\begin{eqnarray}
\label{sol.formA.res}
    A & = & \cos{\frac{\Omega\zeta}{2}} \, ,
   \\
\label{sol.formB.res}
    B & = & i e^{ i\phi_d} \sin{\frac{\Omega\zeta}{2}} \, ,
\end{eqnarray}
where we have used the relations (\ref{mu_1,2})-(\ref{delta}).
Hence, if we use the ansatz (\ref{psi}) and
Eqs.~(\ref{psi_a})-(\ref{psi_b}) and then recover the corresponding
expressions of generic notations $\Omega_R$, $\phi_d$ and $\zeta$,
we will arrive to the form (\ref{solution1}) and (\ref{solution2})
for the laser-driven two-level atom in collinear- and crossed-beam
scenario, respectively.

To summarize this derivation, we have found exact solutions
(\ref{sol.formA.res})-(\ref{sol.formB.res}) for a particular
coupling of the atomic motion to the radiation field by applying the
eikonal, long-wave and rotating-wave approximations. In practice,
these three approximations are well justified if the frequency of
the electric field is nearly resonant to the two-level excitation
energy (for the RWA), the laser wavelength is larger than the atomic
sizes (for the LWA) and since the de' Broglie wavelength of an atom
is typically much smaller than the wavelength of the radiation field
(for the EA). Of course, the RWA might fail if the resonance
condition is not fulfilled to a sufficient degree which may lead to
the standard Bloch shift \cite{Bloch:40} or some generalized
Bloch-Siegert shift \cite{Tuorila:10}.

\end{document}